\documentclass[runningheads]{llncs}
\usepackage[T1]{fontenc}
\usepackage{graphicx}
\usepackage{listings}
\usepackage{amsmath}
\usepackage{amssymb}
\usepackage{graphicx}
\usepackage{bm}
\usepackage{color}
\usepackage{authblk}
\usepackage{booktabs}
\usepackage{multirow}
\usepackage{float}
\usepackage{tikz}
\usepackage{ctable}
\usepackage[colorlinks, citecolor=green]{hyperref}

\usepackage{color, colortbl}
\definecolor{Gray}{gray}{0.9}
\usetikzlibrary{arrows.meta, arrows}
\usepackage{diagbox}
\usepackage{breakcites}
\usepackage{orcidlink}
\usepackage{setspace}
\usepackage{tabularx}
\usepackage{subcaption}
\usepackage[shortcuts]{extdash}

\begin{document}

\title{Reconstruction-free segmentation from undersampled k-space using transformers}
\author{
Yundi Zhang\inst{1,2}\orcidlink{0009-0008-7725-6369} \and
Nil Stolt-Ansó\inst{1,3}\orcidlink{orcid.org/0009-0001-4457-0967} \and
Jiazhen Pan\inst{1,2}\orcidlink{0000-0002-6305-8117}
Wenqi Huang\inst{1,2} \and
Kerstin Hammernik\inst{1,4}\orcidlink{0000-0002-2734-1409} \and
Daniel~Rueckert\inst{1,2,3,4}\orcidlink{0000-0002-5683-5889}
}
\authorrunning{Y. Zhang et al.}
\institute{School of Computation, Information and Technology, Technical University of Munich, Germany
\and School of Medicine, Klinikum Rechts der Isar, Technical University of Munich, Germany
\and Munich Center for Machine Learning, Technical University Munich
\and Department of Computing, Imperial College London, UK
\\
\email{yundi.zhang@tum.de}
}

\maketitle              

\section{Synopsis}
\noindent \underline{\textbf{Motivation:}}
High acceleration factors place a limit on MRI image reconstruction. This limit is extended to segmentation models when treating these as subsequent independent processes.  

\noindent \underline{\textbf{Goal:}}
Our goal is to produce segmentations directly from sparse k-space measurements without the need for intermediate image reconstruction.

\noindent \underline{\textbf{Approach:}}
We employ a transformer architecture to encode global k-space information into latent features. 
The produced latent vectors condition queried coordinates during decoding to generate segmentation class probabilities.

\noindent \underline{\textbf{Results:}} The model is able to produce better segmentations across high acceleration factors than image-based segmentation baselines.

\noindent \underline{\textbf{Impact:}}
Cardiac segmentation directly from undersampled k-space samples circumvents the need for an intermediate image reconstruction step. This allows the potential to assess myocardial structure and function on higher acceleration factors than methods that rely on images as input.

\section{Introduction}
In cardiac magnetic resonance (CMR) imaging, an abundance of quantitative clinical metrics (such as ejection fraction, strain, etc.) are derived from segment\-ation-based modeling of the myocardium. Image reconstruction and segmentation are typically thought of as independent serial processes. In order to reduce acquisition time, k-space data is usually undersampled and reconstruction techniques are employed. These approaches attempt to recover the pixel-level detail lost during this process. However, accurate segmentation does not strictly benefit from this level of precision, often relying on high level information about the overall content of the image. 

While segmentation of cardiac images predominantly takes place on clean images~\cite{bai2018automated}, previous works have attempted to tackle higher accelerations by performing segmentation directly on unrefined images~\cite{schlemper2018cardiac}. Formulating the task as an end-to-end learning problem has shown further improvements~\cite{frnet}.

We hypothesize that the process of magnetic resonance (MR) image reconstruction requires larger amounts k-space samples than what theoretically would be required to extract a segmentation signal from the raw data. Under this assumption, direct segmentation from k-space has the potential to allow the quantification of relevant clinical metrics under higher acceleration factors, while further decreasing acquisition time. 

\section{Method}
In this work, we demonstrate that Transformers~\cite{vaswani2017attention} are capable of employing global attention to leverage all available k-space measurements to predict accurate segmentation maps. The architecture is able to perform this task directly from a set of sampled k-space points, without a need for zero-filling or interpolating the k-space, and without any form of intermediate reconstruction step. We postulate that multi-headed attention, unlike convolutional approaches, offers the necessary properties to appropriately process the nature of k-space: (1) the mechanism considers global correlations, (2) feature extraction should be insensitive to the relative order in which the same samples are presented, (3) inputs of arbitrary sparsity are supported.

\begin{figure*}[ht]
    \includegraphics[width=\textwidth]{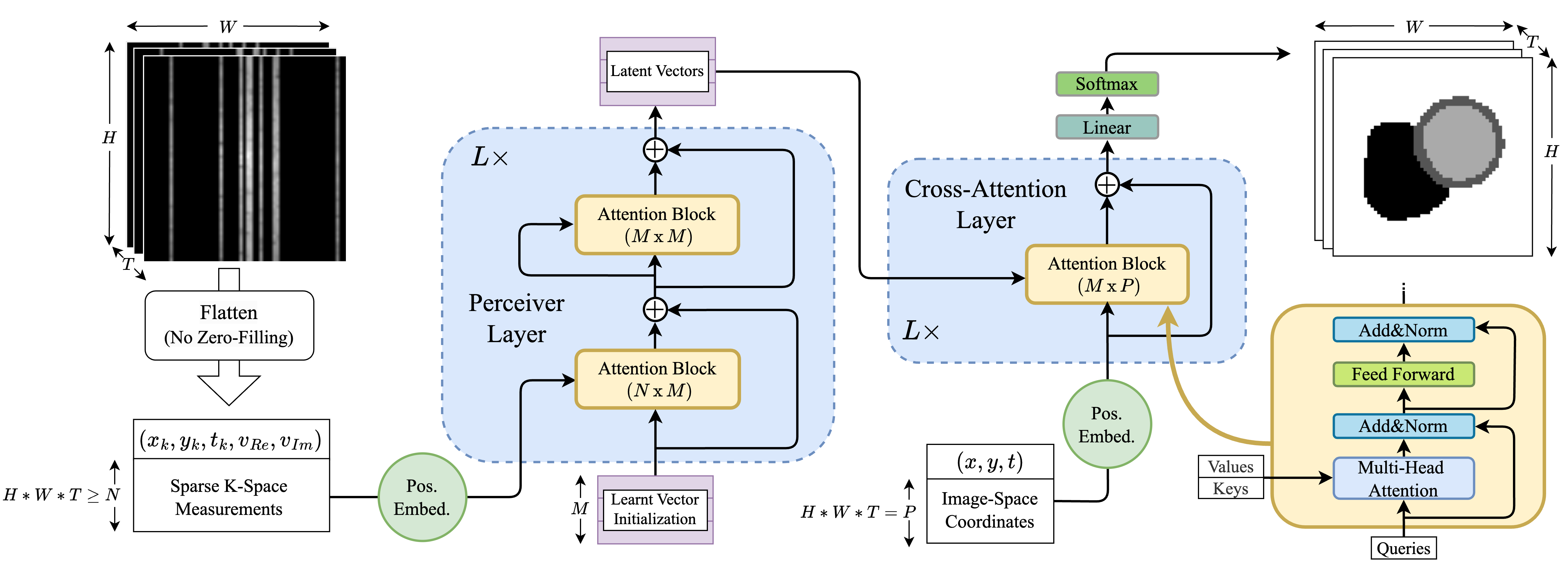}
    \centering
    \caption{Overview of DiSK architecture. $N$, $M$, and $P$ represent the number of sampled points in k-space, latent vectors, and query points in image domain, respectively. The encoder is made up of $L$ Perceiver layers~\cite{Jaegle2021} which alternates cross-attention and latent self-attention blocks. Specifically, cross-attention blocks project global features of the set of input k-space samples $K$ into a fixed-dimensional latent bottleneck of $M$ vectors. Self-attention between latent vectors contextualizes the extracted global features between the vectors. The decoder is made up of $L$ cross-attention layers which condition queried image-domain coordinates with the encoder's latent vectors into segmentation class probabilities.}
    \label{fig:architecture}
\end{figure*}

An overview of the architecture is presented in Fig.~\ref{fig:architecture}. Our architecture's encoder extracts features from the sparse input k-space samples into a latent space over the course of 4 layers. In order to efficiently handle hundreds of thousands of k-space samples while avoiding the $\\mathcal{O}\left(N^2\right)$\ memory complexity of naive self-attention in standard transformers, our encoder utilizes alternating cross-attention (CA) and self-attention (SA) blocks as proposed by Perceiver~\cite{Jaegle2021}. The CA blocks project global k-space information into a fixed bottleneck of latent vectors, while the SA blocks contextualize features between latent vectors.

The decoder consists of 4 cross-attention blocks, which use the extracted latent information to condition any queried image-domain coordinate into producing segmentation class probabilities. The segmentation output is supervised on Dice and binary cross-entropy losses. 

\section{Implementation and Results}
Our training dataset is comprised of 1200 mid-ventricular slices of cardiac short-axis scans from the UK-Biobank dataset~\cite{petersen2015uk}. The dataset is divided into training, validation, and testing sets containing 1000, 100, and 100 samples, respectively. Each scan has 50 frames with an average in-plane resolution of approximately 80x80 pixels per frame. We create synthetic undersampled k-space data on-the-fly for each 2D+time scan. Each frame, we apply additional Gaussian B0 variations (in order to remove the conjugate symmetry of k-space) and generate Cartesian undersampling masks by sampling normally distributed lines centered on the DC component. Implementation and training was performed using the Pytorch library on an NVIDIA A40 GPU.

\begin{figure}[h]
\centering
    \resizebox{.8\textwidth}{!}{%
    \begin{tikzpicture}[font=\LARGE]
    \node[anchor=north](Acc.) at (-0.3,14.4) {Acc.};
    \node[anchor=north](GT) at (1.6,14.4) {GT};
    \node[anchor=north](Syn-Net) at (4.7,14.4) {Syn-Net};
    \node[anchor=north](Li-Net) at (7.8,14.4) {LI-Net};
    \node[anchor=north](Ours) at (10.7,14.4) {Ours};
    \node[anchor=north](k-space) at (13.6,14.4) {Zero-filling};
    \node[anchor=east](4x) at (0.2,12.26) {$4 \times$};
    \node[anchor=south west, scale=1.06] (4x/gt) at (0,10.96)
    {\includegraphics[width=2.8cm]{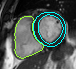}};
    \node[anchor=south west, scale=1.06] (4x/unet) at (3,10.96)
    {\includegraphics[width=2.8cm]{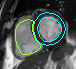}};
    \node[anchor=south west, scale=1.06] (4x/linet) at (6,10.96)
    {\includegraphics[width=2.8cm]{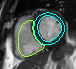}};
    \node[anchor=south west, scale=1.06] (4x/ours) at (9,10.96)
    {\includegraphics[width=2.8cm]{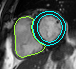}};
    \node[anchor=south west, scale=1.06] (4x/kspace) at (12,10.96)
    {\includegraphics[width=2.8cm]{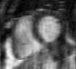}};
    \node[anchor=east](8x) at (0.2,9.52) {$8 \times$};
    \node[anchor=south west, scale=1.06] (8x/gt) at (0,8.22)
    {\includegraphics[width=2.8cm]{Figures/acc/gt.png}};
    \node[anchor=south west, scale=1.06] (8x/unet) at (3,8.22)
    {\includegraphics[width=2.8cm]{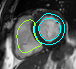}};
    \node[anchor=south west, scale=1.06] (8x/linet) at (6,8.22)
    {\includegraphics[width=2.8cm]{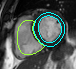}};
    \node[anchor=south west, scale=1.06] (8x/ours) at (9,8.22)
    {\includegraphics[width=2.8cm]{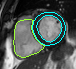}};
    \node[anchor=south west, scale=1.06] (8x/kspace) at (12,8.22)
    {\includegraphics[width=2.8cm]{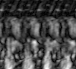}};
    \node[anchor=east](16x) at (0.2,6.78) {$16 \times$};
    \node[anchor=south west, scale=1.06] (16x/gt) at (0,5.48)
    {\includegraphics[width=2.8cm]{Figures/acc/gt.png}};
    \node[anchor=south west, scale=1.06] (16x/unet) at (3,5.48)
    {\includegraphics[width=2.8cm]{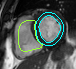}};
    \node[anchor=south west, scale=1.06] (16x/linet) at (6,5.48)
    {\includegraphics[width=2.8cm]{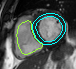}};
    \node[anchor=south west, scale=1.06] (16x/ours) at (9,5.48)
    {\includegraphics[width=2.8cm]{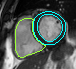}};
    \node[anchor=south west, scale=1.06] (16x/kspace) at (12,5.48)
    {\includegraphics[width=2.8cm]{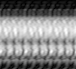}};
    \node[anchor=east](32x) at (0.2,4.04) {$32 \times$};
    \node[anchor=south west, scale=1.06] (32x/gt) at (0, 2.74)
    {\includegraphics[width=2.8cm]{Figures/acc/gt.png}};
    \node[anchor=south west, scale=1.06] (32x/unet) at (3,2.74)
    {\includegraphics[width=2.8cm]{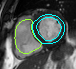}};
    \node[anchor=south west, scale=1.06] (32x/linet) at (6,2.74)
    {\includegraphics[width=2.8cm]{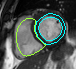}};
    \node[anchor=south west, scale=1.06] (32x/ours) at (9,2.74)
    {\includegraphics[width=2.8cm]{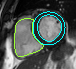}};
    \node[anchor=south west, scale=1.06] (32x/kspace) at (12,2.74)
    {\includegraphics[width=2.8cm]{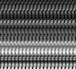}};
    
    \node[anchor=east](64x) at (0.2,1.3) {$64 \times$};
    \node[anchor=south west, scale=1.06] (64x/gt) at (0,0)
    {\includegraphics[width=2.8cm]{Figures/acc/gt.png}};
    \node[anchor=south west, scale=1.06] (64x/unet) at (3,0)
    {\includegraphics[width=2.8cm]{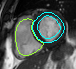}};
    \node[anchor=south west, scale=1.06] (64x/linet) at (6,0)
    {\includegraphics[width=2.8cm]{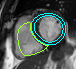}};
    \node[anchor=south west, scale=1.06] (64x/ours) at (9,0)
    {\includegraphics[width=2.8cm]{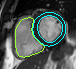}};
    \node[anchor=south west, scale=1.06] (64x/kspace) at (12,0)
    {\includegraphics[width=2.8cm]{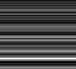}};
    
    \end{tikzpicture}
    }
    \caption{Test set segmentations predicted from different models over varying acceleration factors. The last column visualizes the undersampled k-space measurements in a time frame.}
    \label{fig:acc}
    \end{figure}

We evaluate the performance of models trained on acceleration factors 4, 8, 16, 32, and 64. Our approach is compared to the performance of two image-based segmentation baselines. Following~\cite{schlemper2018cardiac}, we implement a model based on the U-Net~\cite{ronneberger2015u} architecture (Syn-Net) and an autoencoder approach (LI-Net). Their work showed these models to be capable of producing segmentations on noisy reconstructions of undersampled images. As shown in Tab.~\ref{table:scores} and Fig.~\ref{fig:acc}, our model obtains higher segmentation Dice scores and lower Hausdorff distances than the proposed baselines across all tested accelerations.
\begin{table}[h]
    \centering
    \caption{Dice scores and Hausdorff distances over a testing set of 100 subjects.}
    \resizebox{0.7\columnwidth}{!}{
    \begin{tabular}{c c c c c}
    \toprule \multirow{2}{*}{  } Acc. &   & Syn-Net & LI-Net &  Ours \\
    \hline \rowcolor[gray]{0.9}  & Dice $\uparrow$ & $0.749_{ \pm 0.260}$ & $0.805_{ \pm 0.198}$ & $\bm{0.902_{ \pm 0.089}}$ \\
    \rowcolor[gray]{0.9} \multirow{-2}{*}{4$\times$} & HD \: $\downarrow$& $6.809_{ \pm 2.776}$ & $6.557_{ \pm 3.160}$ & $\bm{4.797_{ \pm 2.064}}$ \\
    \hline & Dice $\uparrow$ & $0.748_{ \pm 0.258}$ & $0.809_{ \pm 0.192}$ & $\bm{0.902_{ \pm 0.089}}$ \\
    \multirow{-2}{*}{8$\times$} & HD \: $\downarrow$ & $6.794_{ \pm 2.868}$ & $7.019_{ \pm 3.521}$ & $\bm{4.772_{ \pm 2.253}}$ \\
    \hline \rowcolor[gray]{0.9}  & Dice $\uparrow$ & $0.742_{ \pm 0.264}$ & $0.800_{ \pm 0.197}$ & $\bm{0.903_{ \pm 0.085}}$ \\
    \rowcolor[gray]{0.9} \multirow{-2}{*}{16$\times$} & HD \: $\downarrow$ & $6.792_{ \pm 2.818}$ & $6.841_{ \pm 2.971}$ & $\bm{4.509_{ \pm 2.068}}$ \\
    \hline & Dice $\uparrow$ & $ 0.723_{ \pm 0.287}$ & $0.752_{ \pm 0.242}$ & $\bm{0.902_{ \pm 0.086}}$ \\
    \multirow{-2}{*}{32$\times$} & HD \: $\downarrow$ & $7.383_{ \pm 3.122}$ & $7.531_{ \pm 3.131}$ & $\bm{4.665_{ \pm 2.054}}$ \\
    \hline \rowcolor[gray]{0.9}  & Dice $\uparrow$ & $0.733_{ \pm 0.261}$ & $0.799_{ \pm 0.190}$ & $\bm{0.902_{ \pm 0.085}}$ \\
    \rowcolor[gray]{0.9} \multirow{-2}{*}{64$\times$} & HD \: $\downarrow$ & $7.543_{ \pm 2.972}$ & $6.706_{ \pm 2.567}$ & $\bm{4.911_{ \pm 2.356}}$ \\
    \bottomrule
    \end{tabular}
    }
    \label{table:scores}
\end{table}  

\section{Discussion}
Due to the nature of short-axis acquisitions, the heart is consistently on the same general location and orientation across the dataset. It is therefore easy for the models to achieve a decent performance at high accelerations by memorizing a general shape and location. Despite this, our model appears to consistently produce better approximations of the true anatomy. We suspect that our model's ability to attend globally across all time frames plays a key role. 

\section{Conclusion}
To the best of our knowledge, this is the first study that explores the prediction of cardiac segmentation maps directly from sparse under-sampled k-space measurements without an explicit image reconstruction step. Our results show that transformer architectures are capable of extracting global features from sparse k-space measurements and improve segmentation performance over image-based baselines at high acceleration factors.

\bibliographystyle{splncs04}
\bibliography{refs.bib}

\end{document}